# Expectations from LHC and LC on Top Physics

Marina Cobal

*University of Udine and INFN Trieste, Italy*

**Abstract.** The current status of top quark studies at the Large Hadron Collider (LHC) and at the Linear Collider (LC) is summarized. Expectations on the measurements of the top mass, couplings and decays, of the top spin polarization and of the single top production are here reviewed.

## INTRODUCTION

The present level of understanding of the fundamental forces and of the matter constituents (described by the so-called Standard Model -SM) will improve by the experiments under construction at the Large Hadron Collider (LHC) [1] and those planned for the International Linear Collider (ILC) [2]. The LHC, which will collide protons with protons at a center-of-mass energy of 14 TeV is currently under construction and will go into operation in 2007. The ILC, which will bring the electron to collision with its antiparticle, has been agreed in a world-wide consensus to be the next large experimental facility in high-energy physics. The LHC and ILC will be characterized by distinct experimental conditions. The LHC, due to its high collision energy, in particular has a large mass reach for direct discoveries of new physics. Striking features of the ILC are its clean experimental environment, polarised beams, and known collision energy, enabling precision measurements and therefore detailed studies of directly accessible new particles as well as a high sensitivity to indirect effects of new physics

The physics of the top quark plays an important role as a possible window to new physics. The top quark is the heaviest elementary particle found so far. Since it decays much faster than the typical time for formation of top hadrons, it provides a clean source of fundamental information. Accurate measurements of the top quark properties, such as its mass, couplings, in particular the couplings to gauge bosons and Higgs fields, and branching ratios of rare decay modes, probe possible deviations from the SM predictions in a sensitive way. Since there are many possibilities of anomalous couplings of the top quark to gauge bosons, one will greatly benefit from the combined results of the LHC and LC. Also from the experimental point of view top quarks are important. Since it will be abundantly produced at LHC, it will provide an essential tool for understanding the detectors during the commissioning phase. Triggering, tracking, *b*-tagging, energy and jet calibration can all benefit from the present top signal. Lastly, top quarks will be a major source of background for almost all searches for physics beyond the SM. Precise understanding of the top signal is crucial to claim new physics. The physics of the top quark at the LHC and LC has been studied in great detail, including in some cases a realistic simulation of the detectors. Here I try to summarize the basic predictions.

## TOP-QUARK PRODUCTION AND DECAY

At hadron and lepton colliders, top quarks may be produced either in pairs or singly. The pair-production cross section, about 850 pb, is known at the LHC to NLO level [3] including the re-summation of the Sudakov logarithms (NLL) [4]. About 90% of the rate is due to gluon-gluon collisions, while quark-antiquark collisions give the remaining 10%. The estimated overall theoretical systematic uncertainty is about 12%, At the LHC more than 8 million pairs of top per year at low luminosiy. The LHC will be therefore a "top quark factory", allowing a very large variety of top physics studies with the high statistics samples that will be accumulated. The total top-quark pair production cross section at the LC in the continuum has been calculated to the NLO QCD [5] and 1-loop EW [6] level. The cross section is about $10^3$ times smaller than at the LHC approaching about 0.85 pb at maximum around 390 GeV and falling down with the energy as $1/s$.

## TOP MASS

The mass of the top quark $m_t$ is a fundamental parameter of the SM. Also, in the SM, $m_t$ is related to the masses of the W and Higgs bsons through radiative corrections. Then, the more accurately $m_t$ is known, the tighter constraints can be put on the Higgs mass.

At the LHC, the inclusive lepton plus jet channel, $t\bar{t} \to W^+W^-b\bar{b} \to l\nu jjb\bar{b}$ provides a large and clean sample of top quarks and is the most promising channel for an accurate determination of $m_t$ [7]. Considering only electrons and muons, the BR of this channel is 29.6%. Already in the commissioning phase of the detectors, during the startup of LHC, it will be possible to observe a top signal. ATLAS performed a study [8] in which $b$-jet tagging is assumed to be absent, as pessimistic scenario. In this case $m_t$ is reconstructed from a sample of events with exactly four reconstructed jets and the three jets that result in the highest $P_T$ are used to calculate the invariant mass. After 150 pb$^{-1}$ (few days of running) a clear top peak will be visible already

Various methods have been exploited to measure precisely $m_t$. In the most straightforward method the hadronic decay is used, and the top mass is obtained from the invariant mass of the three jets coming from the same top: $m_t = m_{jjb}$

The typical selection of single lepton top events is based on the presence of an isolated high $P_T$ lepton with $P_T > 20$ GeV and missing energy $E_T^{miss} > 20$ GeV.

At least four jets, reconstructed with a cone size of $\Delta R = 0.4$, with $P_T > 40$ GeV and $|\eta| < 2.5$ are required. One or two jets are required to be tagged as $b$-jets. The reconstruction of the decay $W \to jj$ is first performed by selecting the pair of non $b$- tagged jets with invariant mass closest to $m_W$. Events are retained if $|m_{jj} - m_W| < 20$ GeV. The combination of the jet pair $jj$ with the $b$-tagged jet yields a combinatoric ambiguity. For events with only one tagged $b$-jet the events are kept for which the opening angle of the $b$-jet with the $W$ is smaller than with the lepton of the event. For events with two $b$-tagged jets, the $b$-jet which resulted in the highest $P_T$ of the system was combined with the jet pair $jj$. In Figure 1 (a) the reconstructed $m_t$ for the ATLAS detector is shown. The statistical uncertainty on $m_t$ is not a problem and the background is well under control, with a signal to background ratio S/B~65.

The largest systematic uncertainties arise from the jet energy scale, the $b$-quark fragmentation, the initial and final state radiation and the background contributions. The studies indicate that a total error on $m_t$ below 2 GeV should be feasible, possibly reaching an ultimate precision around 1 GeV. The precision achieved by the LHC will determine the position and size of the interval for the energy scan at the LC

At the LC in fact, the mass will be found mainly from the threshold behavior of the cross section (Figure 1 b)). The threshold measurement implies a comparison data-Monte Carlo, and involves a transition from the actually measured quantity to a suitably defined (short distance) $m_t$, like the MS mass. Several NNLO calculations have been performed for various definitions of the threshold top-quark mass parameter [9]. The large top width provides IR cutoff, so one can use perturbative QCD to compute the cross section. Converge is sensitive to the mass definition used: the pole and kinematic masses are not IR-safe. The best results come from using the 1s mass definition (1/2 the mass of the lowest tt bound state, evaluated in the limit $\Gamma_t \to 0$) combined with a velocity resummation. A delicate comparison of different calculations leads to the conclusion that a scan at the threshold is expected to reduce the error on the top mass down to $dm_t £ 100$ MeV [10] (mainly due to the theoretical uncertainties to relate the 1s with the MS mass), a value not achievable at hadron machines.

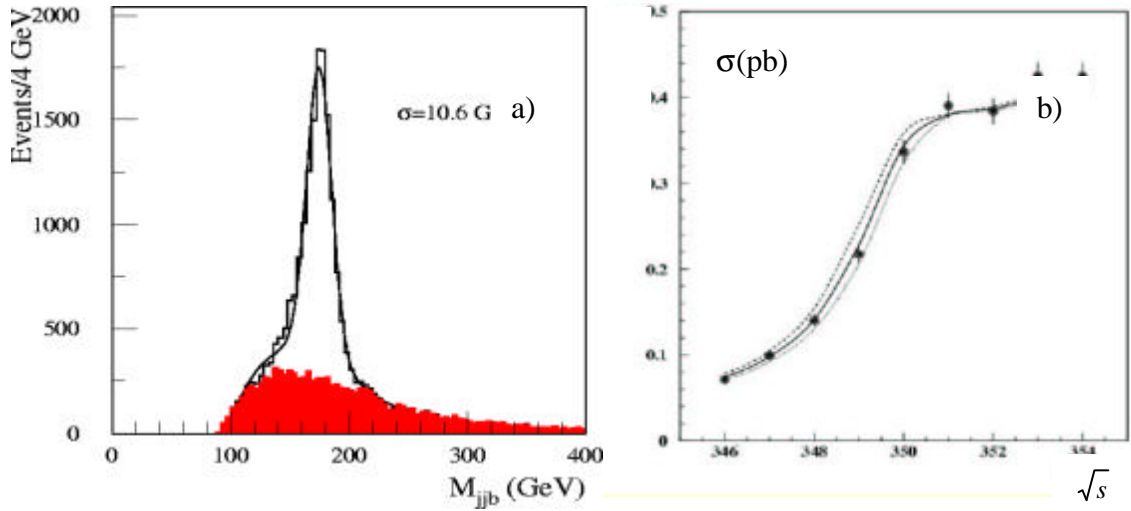

**FIGURE 1.** a) Reconstructed $m_t$ at the ATLAS experiment for 150 pb$^{-1}$ of data. The background corresponds to $W$+4 jet events. b) Top cross section as a function of the cms energy at the LC (treshold scan).

# SEARCH FOR RESONANCES

Physics beyond the SM could affect cross section measurements for $t\bar{t}$ production in a variety of ways (as predicted in SM Higgs, MSSM Higgs, Technicolor models, strong electro-weak simmetry breaking models, Topcolor, etc.): a heavy resonance decaying to $t\bar{t}$ might enhance the cross section, and might produce a peak in the $t\bar{t}$ invariant mass spectrum. Because of the large variety of models and their parameters, in ATLAS (LHC) a study was made[1] of the sensitivity to a "generic" narrow resonance decaying to $t\bar{t}$. Events of the $t\bar{t} \to WWb\bar{b} \to l\nu jjb\bar{b}$ topology were selected. In addition it was required a number of jets with $p_T > 20 GeV$ and $\eta$<3.2 between four and ten, with at least one of them tagged as $b$-jet. After these cuts, the background is dominated by the $t\bar{t}$ continuum. The hadronic $W \to jj$ decay was reconstructed by selecting pairs of jets from among those not tagged as $b$-jets. In cases where there were at least two $b$-tagged jets, then candidates for $t \to Wb$ were formed by combining the $W \to l\nu$ and $W \to jj$ candidates with each of them. In events with only a single $b$-tagged jet, this was assigned as one of the $b$-quarks and each of the still unassigned jets was considered as a candidate for the other $b$-quark. Among the many different possible combinations, the one was chosen which minimized a $\chi^2$. The obtained mass resolution $\sigma(m_{tt})/m_{tt}$ was equal to 6.6%. As an example, Figure 2 a) shows the reconstructed $m_{tt}$ distribution for a narrow resonance of mass 1600 GeV.

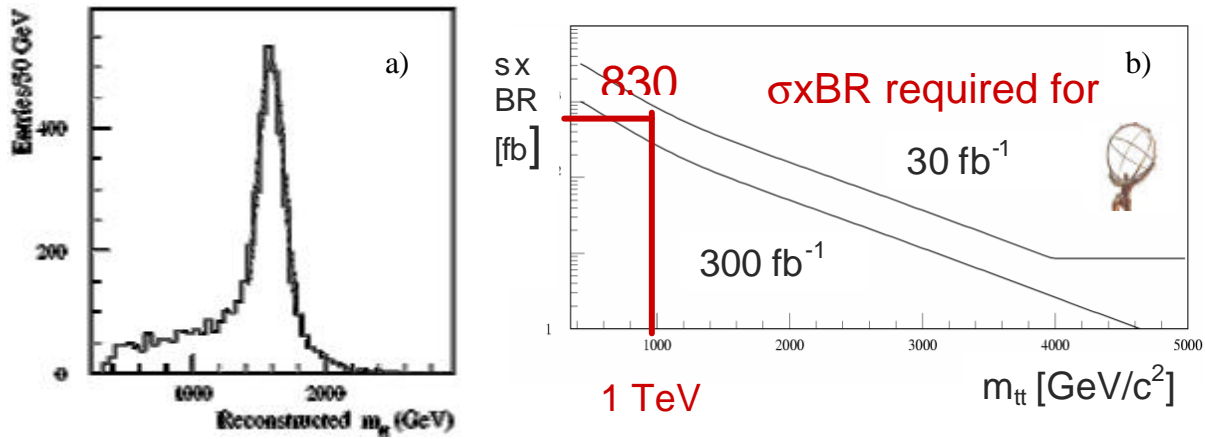

**FIGURE 2.** a) Measured invariant mass distribution for reconstruction of a narrow resonance of mass 1600 GeV decaying to $t\bar{t}$. b) Value of $\sigma$xBR required for a $5\sigma$ discovery potential for a narrow resonance decaying to $t\bar{t}$ as a function of $m_{tt}$.

The reconstruction efficiency, not including BRs, was about 20% for a resonance of mass 400 GeV, decreasing gradually to about 15\% for $m_t = 2$ TeV. For a narrow resonance $X$. Figure 2 b) shows the required $\sigma$xBR( $X \to t\bar{t}$ ) for discovery (the signal must have a statistical significance of at least $5\sigma$ and must contain at least 10 events). Results are shown as a function of $m_X$ for 30 fb$^{-1}$ and 300 fb$^{-1}$.

# COUPLINGS AND DECAYS

Within the SM the top quark decays via a pure V-A interaction, and the decay $t \to Wb$ is dominant with a BR~99.9%. Expectations for the CKM-suppressed decays are approximately 0.1% and 0.01% for $t \to Ws$ and $t \to Wd$, respectively. Determining whether the top quark has the couplings and decays predicted by the SM provides a sensitive probe of physics beyond the SM. For example, in ATLAS the possibility to determine the top quark charge has been analyzed using two different approaches: the reconstruction via its decay products and by determination of photon-top quark coupling in the radiative $t\bar{t}$ events. In addition, studies have been done to check that the $t\bar{t}$ spin correlation predicted in the SM can be observed, and used to probe for anomalous coupling or CP violation. Another interesting point is that in the SM the mass of the top quark is due to its Yukawa coupling $y_t$ to the Higgs boson. The value of $y_t$ can be accessed experimentally by searching for $t\bar{t}H$ production. This analysis requires the $t\bar{t}$ pair to decay semi-leptonically, and since the cross section production is relevant only for a light Higgs, the Higgs boson is detected through its decay $H \to b\bar{b}$, the dominant decay channel for the $m_H$ range of interest. For a $m_H$=120 GeV, 40 (62) signal events with a background of 127 (257) events are expected with 30 (100) fb$^{-1}$, and $y_t$ could therefore be measured with a statistical error of 16.2% for 30 fb$^{-1}$, improving to 14.4% for an integrated luminosity of 100 fb$^{-1}$. Many of the systematic errors could be controlled by comparing the $t\bar{t}H$ rate with the $t\bar{t}$ rate.

## Search for Anomalous Wtb Couplings

The *Wtb* vertex structure can be probed using either top quak pair or single-top quark production processes. The tt cross section depends weakly from it, but there are several sensitive observables, like C and P asymmetries, top quark polarization, and spin correlations which can provide interesting informations. The single top production rate is instead directly proportional to the square of the *Wtb* coupling. The comparison between the LHC and LC potentials is shown in Table 1, having defined: $F_{2L} = \frac{2M_W}{\Lambda} h^W (-f^W - ih^W), F_{2R} = \frac{2M_W}{\Lambda} h^W (-f^W + ih^W)$. One may conclude that the upgraded Tevatron will be able to perform the first direct measurements of the structure of the *Wtb* coupling. The LHC with 5% systematic uncertainties will improve the Tevatron limits considerably, rivaling the reach of a high-luminosity (500 fb$^{-1}$) 500 GeV LC option. A very high energy LC with 500 fb$^{-1}$ luminosity will eventually improve the LHC limits by a factor of three to eight, depending on the coupling under consideration.

**TABLE 1. Uncorrelated limits on anomalous couplings from different machines.**

|  | $F_{L2}$ | $F_{R2}$ |
| --- | --- | --- |
| Tevatron ($\Delta_{sys}$ ~10%) | -0.18…+0.55 | -0.24…+0.25 |
| LHC ($\Delta_{sys}$ ~5%) | -0.052…+0.097 | -0.12…+0.13 |
| $\gamma e$ ($\sqrt{s_{e+e-}} = 0.5 TeV$) | -0.1…+0.1 | -0.1…+0.1 |
| $\gamma e$ ($\sqrt{s_{e+e-}} = 2.0 TeV$) | -0.008…+0.035 | -0.016…+0.016 |

# FCNC Rare Decays

With its large mass, the top quark will couple strongly to the EWSB sector. Many models of physics beyond the SM include a more complicated EWSB sector, with implications for top quark decays. Example includes the possible existence of charged Higgs bosons, or possibly large flavour changing neutral currents FCNC in top decays. The sensitivity to some of these scenario is shown in Table 2.

In the SM, FCNC decays of the top quark are highly suppressed (BR<$10^{-13}$-$10^{-10}$). However, several extensions of the SM can lead to very significant enhancements of these BRs (BR<$10^{-5}$-$10^{-6}$ or even higher).

At present only a few cases (like-sign top-pair production, $t \to qZ$ and $t \to q\gamma$ decays, have been investigated with a more or less realistic detector simulation.

For the LHC case all three possible FCNC decays have been investigated: $t \to q V$, where $V = \gamma, Z, g$ and $q = u$ or $c$. At a linear $e^+e^-$ collider both methods of searching for FCNC interactions were considered (see [11, 12]): $t\bar{t}$-pair production via SM $\gamma/Z$-exchange with following FCNC decays $t \to q\gamma (Z)$: $e^+e^- \to t\bar{t}$, with $t \to q\gamma (Z)$, and single-top-quark production due to FCNC anomalous interactions with the SM decay channel $t \to bW$: $e^+e^- \to tu(c)$, with $t \to bW$.

Note that, due to the limited statistics which could be gathered in a future linear collider in the reaction $e^+e^- \to t\bar{t}$, the LHC has an advantage in the searches for rare top-quark decays.

On the other hand, the future LC has a much smaller background. Therefore, both the LHC and a future LC have great potential to discover top-quark production due to anomalous interactions. Only for anomalous interactions with a gluon (*tgc* or *tgu*) will the LHC have an evident advantage.

**TABLE 2.** Sensitivity to BR for various processes at different machines

| t→ | Tevatron Run II | LHC decay | LHC production | e+e- $\sqrt{s} > 500 GeV$ |
|---|---|---|---|---|
| gq | 0.06% | $1.6 \times 10^{-3}$ | $1 \times 10^{-5}$ | - |
| γq | 0.28% | $2.5 \times 10^{-5}$ | $3 \times 10^{-6}$ | $4 \times 10^{-6}$ |
| Zq | 1.3% | $1.6 \times 10^{-4}$ | $1 \times 10^{-4}$ | $2 \times 10^{-4}$ |

# Top Charge Determination

In order to confirm that its electric charge is indeed $Q_{top}=2/3$, one can either measure the charge of the *b*-jet and the *W*- boson, or attempt to directly measure the top quark coupling through photon radiation in $pp \to t\bar{t}\gamma$ and $pp \to t\bar{t}$ with $t \to Wb\gamma$ [13]. Since the first process is dominated by gluon-gluon fusion at LHC, one expects that the $tt\gamma$ cross-section is approximately proportional to $Q_{top}^2$. The treatment of the radiative top production and top decay matrix elements, fed into the PYTHIA Monte Carlo, was based on the on-mass approach for the decaying top, i.e. the production and decay were treated independently. By suitable selection criteria the hard γ radiation from top production can be enhanced. It is foreseen to disentangle the top charge during the first year of running at LHC using this method.

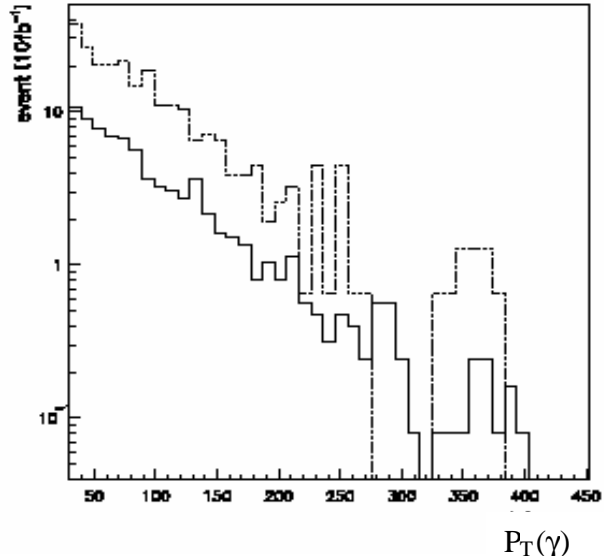

**FIGURE 3.** Yield of hard photon radiation, as a function of the photon $p_T$. The plot corresponds to 10 fb$^{-1}$ of data. The full line corresponds to $Q_{top}=2/3$, the dashed line to to $Q_{top}=4/3$.

## Top Spin Correlations

Since the top quark decays before hadronization, its spin properties are not spoiled. Therefore spin correlations in top production and decays is an interesting issue in top-quark physics. The $t\bar{t}$ spin correlations allow the direct measurement of the top quark spin. In the dileptonic $t\bar{t}$ channel the helicity angles θ* of the leptons are given by the double differential cross section with the asymmetry $\hat{A}$ in the helicity basis defined as the normalized difference between like-spin and unlike-spin pairs. At the LHC where gluon-gluon fusion dominates, the asymmetry is expected to be $\hat{A}$ =0.31± 0.03[14].

Studies are underway to determine also in semi-leptonic top events the top-spin correlations. For top-quark-pair production at LC, and in the quark-antiquark production part at hadron colliders, specifically important at the Tevatron, one can find a top-quark spin-quantization axis, or in other words, a top-spin basis (so called "off-diagonal" basis), in which there will be very strong spin correlations for produced top and anti-top quarks [15]. Detailed studies, both at the $tt$ threshold and in the continuum regions, still remain to be done.

# SINGLE TOP PRODUCTION

The precise determination of the properties of the *Wtb* vertex, and the associated coupling strenghts, will more likely be obtained from measurements of the electroweak production of single top-quarks. At the LHC single top quarks can be produced via three different reactions (shown in Figure 5 a)): *W*-gluon fusion (σ~250*pb*), *Wt* production (σ~60-110*pb*) and *W\** process (σ~10*pb*).

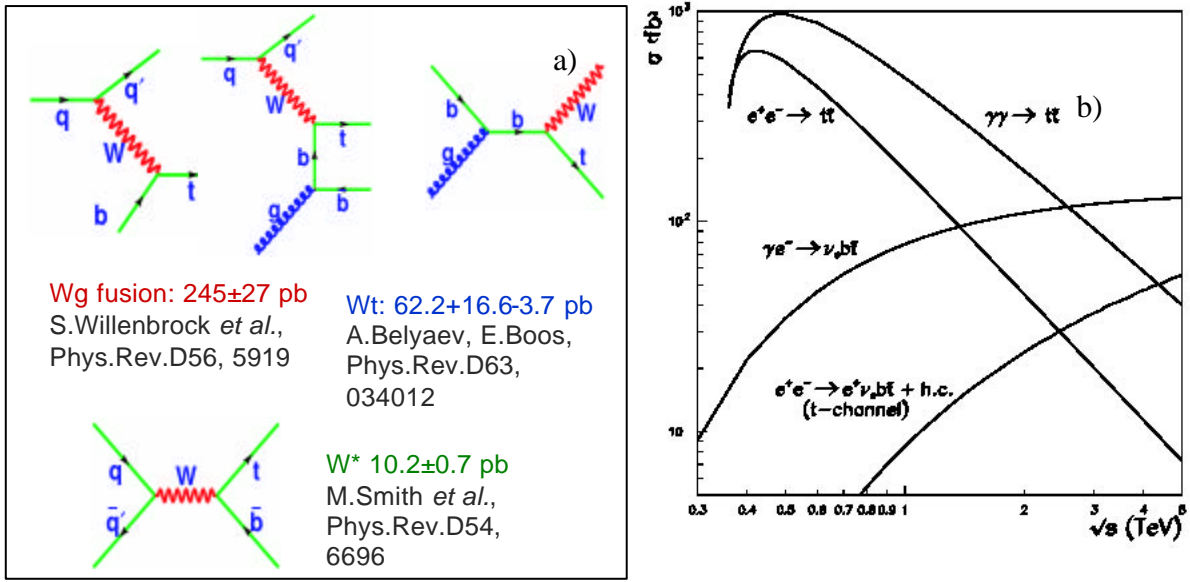

**FIGURE 4.** a) Feyman diagrams for the electroweak single top quark processes accessible at the LHC: 1) *W*-gluon fusion, 2) *Wt* production, 3) s-channel or *W \** process. b) Cross section for top production at the LC.

There are important backgrounds with final states similar to the signals under study (σ(*tt*)=830*pb*, σ(*Wbb*)>300*pb*) and the possibility to extract a signal depends critically from the detector performance. Important parameters are the rate of fake leptons, the *b*-jet identification and fake *b*-jet rate, the capability to identify forward jets and to veto low energy jets.
To reduce the enormous QCD multi-jets background, the $t \to lnj$ decay has been studied, and a pre-selection was done in which a high $P_T$ lepton, at least two jets, at least one *b*-tagged jet, and one forward jet are required. It is really interesting to study the three processes separately, since they have separate sets of backgrounds, their systematic errors for $V_{tb}$ are different, and they are differently sensitive to new physics.
For example, the presence of a heavy *W* would result in an increase of the *W\** signal. Instead, the existence of a *FCNC g⊗t* would be seen in the *W*-gluon fusion channel. Discriminants for the three signals are for example: the jet multiplicity (higher for *Wt*), the presence of more than one jet tagged as a *b* (this increase the *W\** signal with respect to the *W*-gluon fusion one), the mass distribution of the 2-jet system (which has a peak near the *W* mass for the *Wt* signal and not for the others).

At the LC the single-top-quark production cross section was calculated to LO at $e^+e^-$, $\gamma\gamma$ and $\gamma e$ collision modes including various beam polarizations[16]. The recently calculated NLO corrections to the single-top-quark production in $\gamma e$ are well under control and rather small [17]. Single-top-quark production in $\gamma e$ collisions is of special interest; the rate is smaller than the top-pair rate in $e^+e^-$ only by a factor of 1/8 at 500–800 GeV energies, and it becomes the dominant LC processes for top production at a multi-TeV LC like CLIC. Note, the $|V_{tb}|$ matrix element can be measured at a LC significantly more accurately as compared to the LHC. Therefore, its value could be used at LHC in single-top analysis of the process *t*-channel $qb \to q\_t$ to measure the *b*-quark distribution function in the proton.

## Cross Section Measurements

*W-gluon fusion*

The *W-gluon* signal is distinguished from backgrounds by the presence of a spectator-quark jet that emerges in the forward direction. The total number of jets is required to be exactly two, to provide additional rejection of background. Further requirements are that the total mass of the event be greater than 300 GeV and that the sum of the $E_T$ of all the jets and leptons in the event be greater than 200 GeV.

After all the cuts, 27000 events of signal and 8500 of background (mostly *Wjj*) survive, for 30 fb$^{-1}$. This corresponds to a S/B=3.1 and $S/\sqrt{B} = 286$.

The relative statistical uncertainty in the cross section is 0.71%.

*Wt.*

The strategy for measuring the *Wt* cross section is similar to the previous one, since they share the same backgrounds. However, the number of jets in the central region is required to be exactly three (to reduce non-top background). Exactly one of these jets is required to be a *b*-jet. The total invariant mass of all reconstructed leptons and jets was required to be smaller than 300 GeV.

There are 6800 signal events surviving the cuts and 30000 background events (mostly $t\bar{t}$) with a S/B≈0.22 and a $S/\sqrt{B} = 39$. The relative statistical error in the cross section is 2.8%.

*W\**

Since this signal has such a small cross-section, stringent cuts must be made to obtain a reasonable signal-to-background ratio. Exactly two jets are required, with high $p_T$ and tagged as *b*-jets. This cut significantly reduces the *W+jets* background, and also the number of events from the *W-gluon* fusion.

The scalar sum of the $P_T$ of the jets is required to be >175 GeV, and the invariant mass of the event has to be >200 GeV. There are 1100 signal events surviving the cuts and 2400 background events (mostly $t\bar{t}$), with a S/B≈0.46 and a $S/\sqrt{B} = 23$. The relative statistical uncertainty in the cross section is 5.4%. The results are summarized in Table 3.

**TABLE 3.** Relative errors in the cross section measurement.

| Process | Signal | Background | S/B |
|---------|--------|------------|-----|
| Wg fusion | 27k | 8.5k | 3.1 |
| Wt | 6.8k | 30k | 0.22 |
| W* | 1.1k | 2.4k | 0.46 |

The results found for the relative experimental statistical errors on the production cross sections of the single top processes, imply statistical uncertanties on the extraction of $V_{tb}$ of 0.36% for *W-gluon* fusion, 1.4% for *Wt* and 2.7% for *W\**.

The errors in the extraction of $V_{tb}$ would be dominated by uncertainties in the theoretical predictions of the cross-sections. These arise from uncertainties in the parton distribution structure functions (PDFs), uncertainty in the scale ($\mu$) used in the calculation, and the experimental error on the top mass. Table 4 summarizes the errors.

**TABLE 4.** Relative errors in the $V_{tb}$ measurement.

| Process | $dV_{tb}$ (stat) | $dV_{tb}$ (theory) |
|---------|------------------|--------------------|
| Wg fusion | 0.4% | 6% |
| Wt | 1.4% | 6% |
| W* | 2.7% | 6% |

The error due to the uncertainty in the top mass is quoted assuming $\Delta m_t = 2$ GeV. The *Wt* cross section at the LHC is not currently well known theoretically; the value of 50% reported reflects the range of values reported in the literature. The measurement of $V_{tb}$ is also sensitive to errors in the cross-sections of the backgrounds

## COMPARISON BETWEEN LHC AND LC AND CONCLUSIONS

At LHC, $m_t$ will be determined with an ultimate precision of approximately 1 GeV. However, already during the detector commissioning phase the top quark signal will be observed using simple and robust reconstruction methods. LC should then be able to provide a precision of about 100MeV on $m_t$. At the LHC, the high statistics samples available will allow to determine top-quark properties, like charge and spin, unambiguously. Exotic models involving top will be tested with high precision. Limits on rare FCNC top decays will be improved by orders of magnitude, or even observed experimentally. The observation of single top production will provide a precise determination of $V_{tb}$.

Few final remarks can be added comparing LHC and LC: for what concern the kinematics of top events, clearly an $e^+e^-$ linear accelerator is in advantage, since it can makes use of the momentum conservation (while the pp collider can just profit of the $P_T$ conservation). At the LHC the composite nature of the protons will give raise to the underlying events and will make that the center of mass energy is not a fixed number. Moreover, being strongly interacting particles, they will produce a large QCD background. At LC the initial state will be better defined and the background much smaller. However LHC will provide an incredibly large statistics of top events, much higher than at th LC. Given these complementary characteristics, the interplay between LC and LHC should be as useful as it was for LEP+SLC+Tevatron.

Up to short time ago, there was not so much interaction between the LHC and LC communities. However, in 2002 a LHC/LC study group was formed first in Europe and then soon it took a worldwide character. The Working Group contains more than one hundred members from among theorists, physicists from CMS, ATLAS, members of all the LC study groups and Tevatron contact persons. A document is in preparation (the Chapter "Electro-weak and QCD precision physics has a large part dedicated to top physics) and can be found at: www.ippp.dur.ac.uk/~georg/lhclc.